\documentclass[12pt, journal,draftclsnofoot,onecolumn,oneside,a4paper]{IEEEtran}
\usepackage{amsfonts, amssymb}
\usepackage{latexsym}
\usepackage{graphicx}
\usepackage{algorithmic}
\usepackage{color}
\usepackage{setspace}
\usepackage[cmex10]{amsmath}
\usepackage{array}
\usepackage{eqparbox}
\usepackage{stfloats}
\usepackage{mdwmath}
\usepackage{mdwtab}
\usepackage{subfigure}
\usepackage{bm,stmaryrd}
\usepackage{bbm}
\usepackage{multirow}
\usepackage{multicol}
\usepackage{cite}
\newcommand{\indicator}[1]{\mathbbm{1}\left[ {#1} \right]}

\IEEEoverridecommandlockouts

\title{On the Design of a Novel Joint Network-Channel Coding Scheme for the Multiple Access Relay Channel}

\author{Mikel~Hernaez,~\IEEEmembership{Student Member,~IEEE,}
 		Pedro~M.~Crespo,~\IEEEmembership{Senior Member,~IEEE,}
        and Javier~Del~Ser,~\IEEEmembership{Senior Member,~IEEE}%

        \thanks{Mikel Hernaez and Pedro M. Crespo are with Centro De Estudios e Investigaciones T\'ecnicas de Gipuzkoa (CEIT) and TECNUN (University of Navarra), 20008 Donostia-San Sebasti\'an, Spain. E-mails: \{mhernaez, pcrespo\}@ceit.es. Javier Del Ser is with TECNALIA, 48170 Zamudio, Bizkaia, Spain. E-mail: javier.delser@tecnalia.com.}}

\begin{document}
\maketitle
\begin{abstract}
This paper proposes a novel joint non-binary network-channel code for the Time-Division Decode-and-Forward Multiple Access Relay Channel (TD-DF-MARC), where the relay linearly combines -- over a non-binary finite field -- the coded sequences from the source nodes. A method based on an EXIT chart analysis is derived for selecting the best coefficients of the linear combination. Moreover, it is shown that for different setups of the system, different coefficients should be chosen in order to improve the performance. This conclusion contrasts with previous works where a random selection was considered. Monte Carlo simulations show that the proposed scheme outperforms, in terms of its gap to the outage probabilities, the previously published joint network-channel coding approaches. Besides, this gain is achieved by using very short-length codewords, which makes the scheme particularly attractive for low-latency applications.
\end{abstract}


\section{Introduction}\label{sec:intro}
Node cooperation has been widely shown to improve the performance of wireless networks with several terminals by increasing the robustness of the system to channel variations such as deep fades, as well as by enabling significant energy savings. The essence of node cooperation lies in jointly processing the in-network information by all the constituent nodes of the network, which allows improving the spectral and power efficiency of wireless networks and ultimately, attaining the desired diversity-multiplexing tradeoff without requiring additional complexity (e.g. co-located multiterminal MIMO schemes).

A particular example of relay cooperation in multi-terminal networks is the so-called Multiple-Access Relay Channel (MARC). The MARC is a communication scenario where two or more information sources forward data to a single common destination with the help of an intermediate relay \cite{Kramer00}. In this scenario, the relay can work in full \cite{Kramer00,Kramer05} or half duplex mode. For the half duplex mode, the following transmission strategies have been proposed: i) the Constrained MARC (C-MARC \cite{Sankar04}), where the sources transmit during the first time slot and coordinate with the relay during the second time slot by transmitting information; ii) the Orthogonal MARC (O-MARC, see \cite{Sankar07,Sankar11}), where the sources and the relay transmit over two orthogonal channels; and iii) the Time-Division MARC (TD-MARC, see \cite{Hausl05}), where both sources and the relay convey their data by using three orthogonal channels, i.e. for two sources the total transmission time is divided into three time slots, one for each transmitting node. On the other hand, during the last decade the well-known relaying strategies Decode-and-Forward (DF), Compress-and Forward (CF) and Amplify-and-Forward (AF) originally developed for the conventional relay channel \cite{Cover} have been applied to the aforementioned MARC models in a number of contributions (see e.g. \cite{Kramer00,Kramer05,Sankar04,Sankar07, Sankar11} and references therein).

\begin{figure}[t]
\centering
\includegraphics[width=0.9\columnwidth]{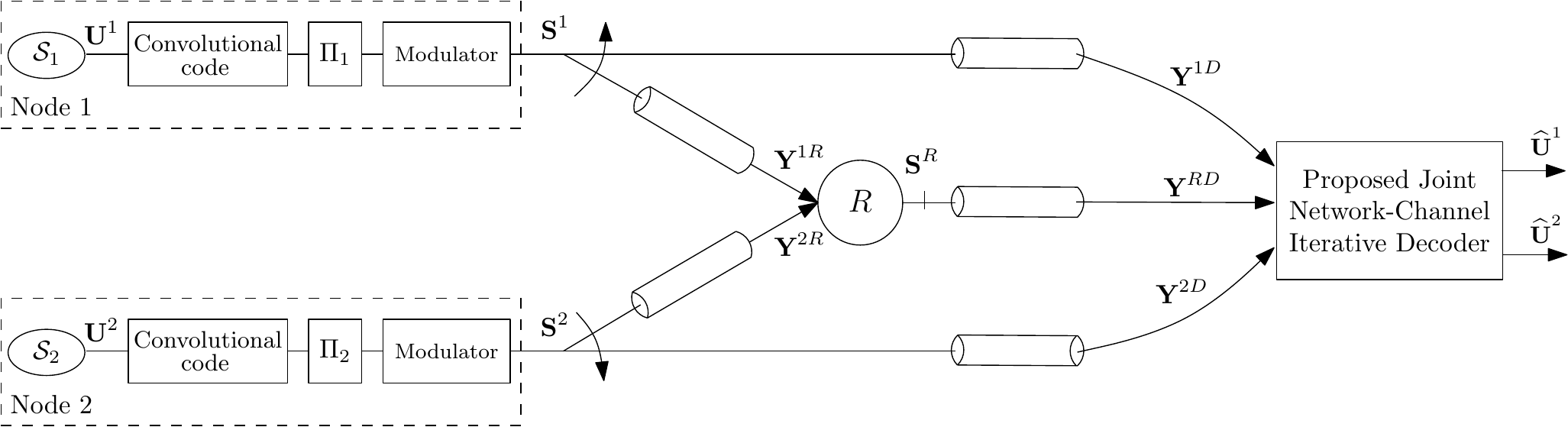}
\caption{Block diagram of the considered $2$-user TD-DF-MARC scenario.}
\label{fig:system}
\end{figure}

When dealing with practical coding schemes for the MARC scenario, the TD-MARC, along with a DF strategy, has been widely studied in the related literature \cite{Hausl05,Hausl06,Nguyen07,Chebli09,Yu10,Ishii11,Yang07,Zeitler09,Xiao09,Guo09,Hernaez11}. Although the time-division scheme involves a suboptimal use of the available bandwidth, it allows for an easier implementation in practical systems thanks to the use of half-duplex relays and the lack of stringent synchronization constraints. Besides, the DF strategy offers a higher code design flexibility. In this work we specifically focus on the 2-user TD-MARC with a DF relaying strategy (hereafter coined as TD-DF-MARC), which is schematically depicted in Figure \ref{fig:system}. It is important to note that the capacity bounds for this model can be derived from the capacity bounds of the O-MARC \cite{Sankar04, Hausl_thesis}.

In the MARC we are interested in maximizing the information conveyed by the relay node. To this end, network coding \cite{Ahlswede00,Li03} has become a widely used technique to complement channel coding schemes used for combating channel-induced errors. Combining the data from both sources at the intermediate relay node embodies a practical tool for approaching the capacity bounds of the TD-DF-MARC scenario. However, by treating network and channel coding separately some performance loss is expected, since the network decoder cannot use the output soft information computed by the channel decodes. Likewise, the channel code cannot exploit the redundant information provided by the network code. This observation is further supported by the findings in \cite{Effros03,Ratnakar06}, where it was shown that in general, capacity can only be achieved by conceiving channel and network coding as a single non-separated data processing stage.

In this context, several practical joint network-channel coding schemes have been recently proposed \cite{Hausl05,Hausl06,Nguyen07,Chebli09,Yu10,Ishii11,Yang07,Zeitler09,Xiao09,Guo09, Gong10, Hernaez11}. In \cite{Hausl05} joint network-channel coding for TD-DF-MARC model was first considered. The authors proposed distributed regular LDPC codes as the joint network-channel code at the relay node, where the destination jointly decodes the messages from the sources with the aid of the information sent from the relay, as opposed to \cite{yChen06} where the two messages transmitted from the sources were separately decoded. In \cite{Hausl06} the authors follow the same idea by proposing a turbo-code-based joint network-channel coding scheme. Parallel to these proposals, the authors in \cite{Nguyen07} proposed a similar scheme for high-order modulations. More recently a joint coding scheme based on WiMax LDPC codes was presented in \cite{Chebli09}, whereas in \cite{Yu10,Ishii11} two schemes based on turbo codes were investigated. Furthermore, some joint non-binary coding schemes have been recently reported in \cite{Xiao09} (non-binary network coding) and \cite{Guo09,Hernaez11} (non-binary network and channel coding). Finally, in \cite{Yang07, Zeitler09} the authors proposed a joint network-channel coding scheme where the relay transmits the soft values resulting from its decoding procedure over AWGN channels.

The present work joins the upsurge of research on the TD-DF-MARC scenario by proposing a novel Joint Network-Channel Code (JNCC) where the relay linearly combines -- over a non-binary finite field -- the coded sequences from the source nodes. The iterative decoding procedure at the common destination is performed by running the Sum-Product Algorithm (SPA \cite{SPA}) on the factor graph describing the proposed JNCC, which is compounded by three sub-factor graphs: two describing the channel codes of each source, and a third describing the network coding operation performed at the relay node. Specifically, the key contributions of this manuscript over the state of the art on this topic are as follows:
\begin{itemize}

\item The proposed scheme does not perform channel coding on the already network-coded bits, reducing the complexity at the relay node without compromising performance. To the knowledge of the authors, all practical schemes for fading channels found in the related literature\footnote{In \cite{Yang07} a similar coding procedure at the relay is proposed; however, their study is restricted to AWGN channels and binary network coding.} perform channel coding on the already network-coded bits.

\item It is shown that a tailored selection of the set of coefficients used in the network coding operation, namely, the Network Coding (NC) coefficients, outperforms a random choice as done in \cite{Guo09}. This selection is performed by matching the EXtrinsic Information Transfer (EXIT, \cite{Brink,EXIT}) functions of the compounding codes on the EXIT chart.

\item Contrary to previous literature, the JNCC allows the sources to use completely different channel codes, at the sole expense of an increased complexity when choosing the NC-coefficients through EXIT-curve matching.

\item The work presented here considers convolutional codes at both source nodes. As a consequence of the previous point, both convolutional codes can be independently terminated. This fact allows us to use very short-length codewords, making the scheme particularly attractive for low-latency applications.
\end{itemize}

The remainder of the manuscript is organized as follows: Section \ref{sec:model} introduces the system model, whereas the  decoding algorithm of the proposed JNCC is detailed in Section \ref{sec:Decoder}. In Section \ref{sec:EXIT} an analysis on the influence of the NC-coefficients is performed through EXIT charts. Section \ref{sec:res} discusses the obtained Monte Carlo simulation results, and finally Section \ref{sec:con} ends the paper by drawing some concluding remarks.

\section{System Model}\label{sec:model}

Let $(\Omega, \beta, \mathcal{P})$ be the underlying probability space where all the random variables (r.v.) are defined. We use uppercase when referring to r.v. and lowercase when referring to realizations of r.v. In addition, we use boldface when referring to vectors; thus, uppercase and boldface refer to random vectors.
For discrete r.v., we denote the probability mass function (p.m.f.) of the discrete r.v. $X$ as $P_{X}(x)\triangleq \mathcal{P}\{X=x\}$. For continuous r.v., we denote the probability density function (p.d.f.) of the continuous r.v. $X$ as $p_{X}(x)$. However, when the context is clear, we use $P(x)$ and $p(x)$ for p.m.f. and p.d.f., respectively.

Referring to Figure \ref{fig:system}, for simplicity we have considered a symmetric scenario consisting of $2$ unit-entropy binary information sources $\mathcal{S}_1$ and $\mathcal{S}_2$, which generate blocks $\mathbf{U}^1\in \{0,1\}^{K}$ and $\mathbf{U}^2\in \{0,1\}^{K}$ of length $K$. As depicted in this figure, at each transmitter the sequence is channel-coded by a convolutional code, producing the codeword $\mathbf{C}^m\triangleq \{\mathbf{C}^m_t\}_{t=1}^N \in \{0,1\}^N$, with $m\in\{1,2\}$ denoting the source index. The code rate is therefore given by $R=K/N$. Each codeword is then interleaved yielding the interleaved codeword $\mathbf{X}^m=\Pi_m\left(\mathbf{C}^m\right)$, where $\Pi_1$ and $\Pi_2$ are two different random spread interleavers with a spread factor equal to $q\in\mathbb{N}$. Finally, the codeword is modulated, resulting in the transmitted sequence $\mathbf{S}^m\triangleq \{S_t^m\}_{t=1}^{N}$, which is transmitted over $M=N/2$ complex dimensions (i.e. $N$ real dimensions). During the first and second time slots source $\mathcal{S}_1$ and $\mathcal{S}_2$ transmit to both the relay and destination nodes the sequence $\mathbf{S}^1$ and $\mathbf{S}^2$, respectively. The third time slot is used by the relay to process the data from the sources and transmit the resulting coded data to destination.

Regarding the links between nodes, we denote as $d_{k,j}$ the distance from transmitter $k\in\{1,2,R\}$ (R: Relay) to receiver $j\in\{R,D\}$ (D:~Destination). Moreover, considering the power at the end of the source-destination link $P_0$ as the reference, the received power at the end of each link will be given by $P_0\cdot(d_{S,D}/d_{k,j})^{\delta}$, where ${\delta}$ denotes an attenuation exponent. In what follows, and without loss of generality, the distances are normalized with respect to $d_{S,D}=1$ and we consider $P_0=1$. Thus, the attenuation undergone by the signals due to the distance-dependant propagation losses of a given link can be expressed as $d_{k,j}^{-\delta}$. Therefore the received symbol per real dimension at each receiver is given by
\begin{equation}\label{eq:io}
Y_t^{m,j}=\alpha^{k,j}\cdot\sqrt{d_{k,j}^{-\delta}}\cdot S_t^k+N_t^{k,j},
\end{equation}
where $\alpha_{k,j}$ is Rayleigh distributed with $E[\alpha_{k,j}^2]=1$ $\forall k,j$, and $\{N_t^{k,j}\}_{t=1}^N$ are modelled as real Gaussian i.i.d. random variables with zero mean and variance $N_0^{k,j}/2$. The values of $\{\alpha_{k,j}\}$ are assumed to remain constant within the duration of a transmitted block (i.e. quasi-static fading). Moreover, full channel state information (CSI) is assumed at the receivers.

\subsection{Relay Node}\label{ssec:relay}

Consider the set of all $2^q$ polynomials $\rho(z)$ of degree $q-1$ with coefficients lying in $GF(2)$ (the binary Galois field). Let $g(z)$ be a prime polynomial (i.e., monic and irreducible polynomial) of order $q$. Then, this set becomes a finite field, $GF(2^q)$, by defining the addition $\oplus$ and multiplication $\otimes$ rules as the $\hspace{-3mm}\mod{g(z)}$ remainder of the sum and product of two polynomials, respectively. Notice that since the $\hspace{-3mm}\mod{g(z)}$ addition rule is just componentwise addition of coefficients in $GF(2)$, $GF(2^q)$ under addition is isomorphic to the vector space $(GF(2))^q$ of binary $q$-tuples with $\hspace{-3mm}\mod{2}$ elementwise addition, denoted hereafter as $\wedge$. Therefore, there is a one-to-one mapping $\psi_q: (GF(2))^q \rightarrow GF(2^q)$ defined as $\psi_q(a_0,\ldots, a_{q-1})= \sum_{k=0}^{q-1} a_{k}z^k$ such that $\psi_q(\mathbf{a}) \oplus \psi_q( \mathbf{b})= \psi_q (\mathbf{a} \wedge \mathbf{b})$, where $\mathbf{a},\mathbf{b}\in (GF(2))^q$. In addition, we index the elements $\rho_i\in GF(2^q)$, $i\in\{0,\ldots,2^q-1\}$ by the base-10 notation of the corresponding binary tuple $(a_0,\ldots, a_{q-1})$. In the following we refer to the elements of the finite field $GF(2^q)$ as \emph{non-binary} symbols.

In the first and second time slots the relay receives the channel sequences $\mathbf{Y}^{1,R}\triangleq\{Y_t^{1,R}\}_{t=1}^{N}$ and $\mathbf{Y}^{2,R}$ from sources $\mathcal{S}^1$ and $\mathcal{S}^2$, respectively and it deinterleaves them. Then, it executes the BCJR algorithm \cite{BCJR} twice in order to obtain the estimations $\widehat{\mathbf{C}}^1$ and $\widehat{\mathbf{C}}^2$ of the source channel-coded sequences, which are then interleaved in order to obtain the estimated interleaved coded bits $\widehat{\mathbf{X}}^1$ and $\widehat{\mathbf{X}}^2$.

Each of the interleaved coded sequences $\widehat{\mathbf{X}}^1$ and $\widehat{\mathbf{X}}^2$ is partitioned into $N/q$ sub-sequences of length $q$. We denote as $V^m=\psi_q(\{\widehat{X}_i^m\}_{i=1}^q)\in GF(2^q)$ to the non-binary symbol associated to the corresponding sub-sequence. The non-binary symbol of the relay $V^R$ is now computed as the linear combination of the non-binary symbols corresponding to each source, i.e.
\begin{equation}\label{eq:lc}
V^R\triangleq\psi_q\left(\psi_q^{-1}\left(h^1 \varotimes V^1\right) \wedge \psi_q^{-1}\left(h^2 \varotimes  V^2\right)\right)\triangleq f(V^1,V^2),
\end{equation}
where $\mathbf{h}=(h^1,h^2)$ and $h^m\in \{\rho_i\}_{i=1}^{2^q-1}$ represents the NC-coefficients used in the linear combination. Finally, the modulated symbols associated to each sub-sequence are computed as $\{S_i^R\}_{i=1}^q=2\cdot \psi_q^{-1}(V^R) - 1$ and the transmitted signal $\mathbf{S}^R$ is obtained by concatenating the $N/q$ resulting modulated sub-sequences.

\section{Iterative Joint Network-Channel Decoder} \label{sec:Decoder}

\begin{figure}[!h]
\centering
\includegraphics[width=0.9\columnwidth]{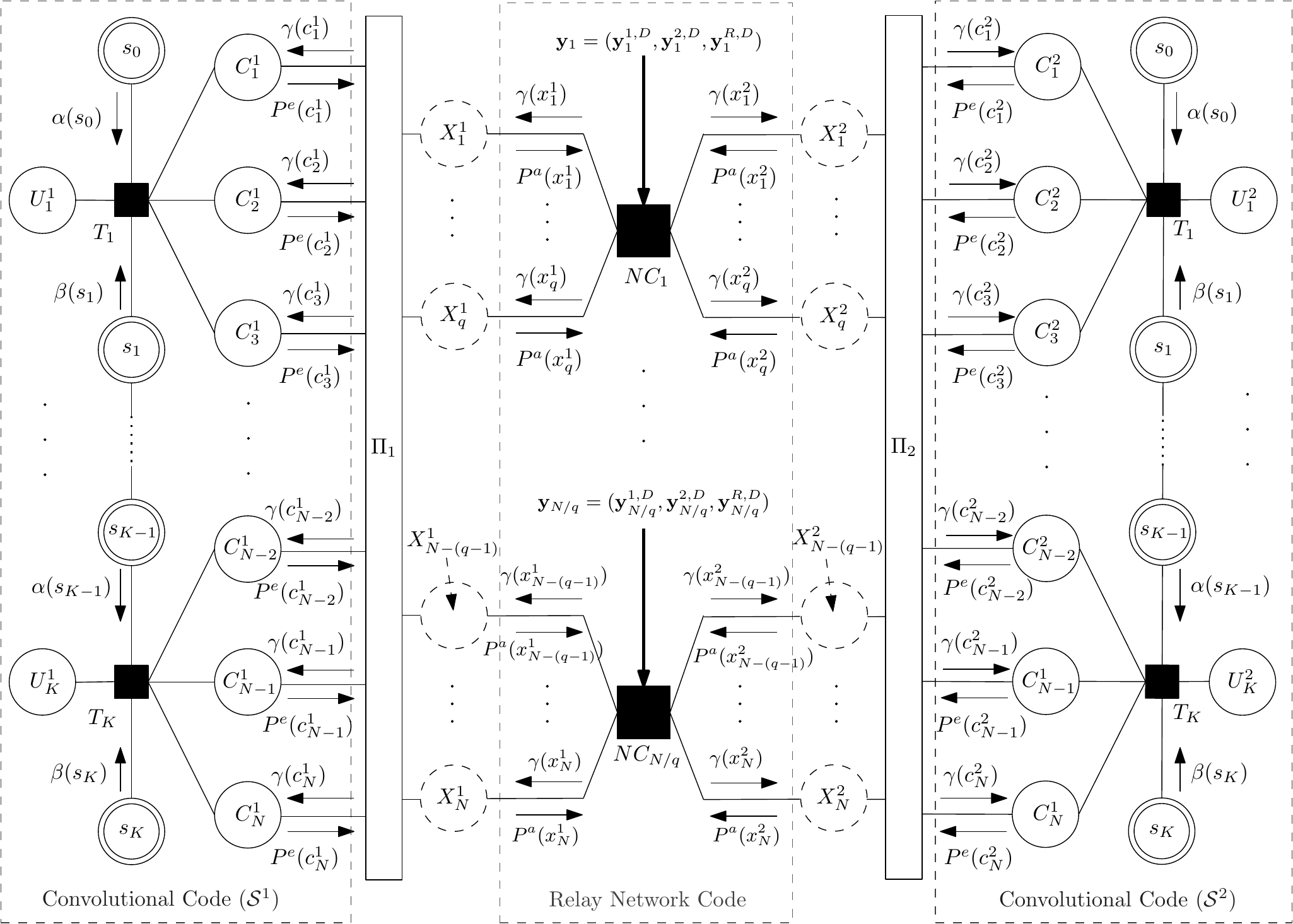}
\caption{Factor Graph of the Proposed JNCC.}
\label{fig:decoder}
\end{figure}

The destination receives the channel outputs $\mathbf{Y}\triangleq(\mathbf{Y}^{1,D},\mathbf{Y}^{2,D},\mathbf{Y}^{R,D})$. The aim of the JNCC decoder is to find the source binary symbols $\{U_k^m\}_{k=1}^K$ that maximize the conditional probability $P(u^m_k|\mathbf{y})$, which is obtained by marginalizing the joint conditional probability $P(\mathbf{u}^m|\mathbf{y})$. This marginalization is efficiently computed by applying the SPA over the factor graph describing $P(\mathbf{u}^m|\mathbf{y})$. Figure \ref{fig:decoder} shows the three compounding sub-factor graphs of the proposed JNCC: two describing the source convolutional codes, and the third one describing the network code used at the relay. As explained in Section \ref{ssec:relay}, the factor graph of the relay network code is in turn composed of  $N/q$ parallel and identical factor nodes, depicted in Figure \ref{fig:decoder} as the oversized factor nodes which we hereafter refer to as \emph{network} check nodes and labelled with $\mathcal{NC}_l$, $l\in\{1,\ldots,N/q\}$. Furthermore, we define $\mathbf{y}_l\triangleq(\mathbf{y}_l^{1,D}, \mathbf{y}_l^{2,D}, \mathbf{y}_l^{R,D})$ as those components of $\mathbf{Y}$ associated to the network check node $\mathcal{NC}_l$. In the next subsection the derivation of the factor graph corresponding to one of these network check nodes is explained, which is then incorporated into the overall factor graph plotted in Figure \ref{fig:decoder}.

Since the overall factor graph of the JNCC has loops, the SPA is iteratively run between the sub-factor graphs corresponding to the relay network code and the convolutional codes. After a fixed number of iterations $\mathcal{I}$, the $U^m_k$ is computed as
\begin{equation}
P(u^m_k|\mathbf{y})\propto\sum_{\sim u^m_k} T_k(s_k,u^m_k,\mathbf{c}_k^m,s_{k+1})\alpha(s_k)\beta(s_{k+1})\prod_{t:\ c_t^m\in \mathbf{c}^m_k} \gamma(c_t^m),
\end{equation}
where $\alpha$ and $\beta$ are the forward and backward messages passed from the adjacent state nodes to the factor node $T_k$ given by the Trellis of the convolutional code; $\mathbf{c}_k^m$ are the coded bits $c_t^m$ associated to $u_k^m$; and $\gamma$ are the messages passed from the variable nodes $c_t^m$ to $T_k$ (i.e. the likelihoods). Note that in the case of not having a relay, the likelihoods $\gamma$ are given by $\gamma(c^m_t)= p(y_t^{m,D}|c_t^m)$. However, when the relay is present, these likelihoods now depend on the messages passed by the network check nodes associated to the interleaved binary symbol $x_{\Pi^{-1}_m(t)}^m$, i.e. $\gamma(c^m_t)=\gamma(x^m_{\Pi^{-1}_m(t)})$, where $\gamma(x^m_t)\propto p(\mathbf{y}_l\vert x_t^m)$. In the next subsection a factorized form of $p(\mathbf{y}_l\vert x_t^m)$ is derived.

\subsection{Factorization of $p(\mathbf{y}_l|x_t^m)$} \label{ssec:NSD}

In this section the factorization of the conditional probability $p(\mathbf{y}_l|x_t^m)$ is derived. As before, the upper index $j$ refers to the random variables associated with both sources and the relay, whereas the index $m$ only refers to variables associated to the sources, i.e. $j\in\{1,2,R\}$ and $m\in\{1,2\}$. Besides we use index $\overline{m}$ to refer to the complement of $m$, i.e. $\overline{m}=3-m$. As $p(\mathbf{y}_l\vert x_t^m)$ only depends on the bits belonging to the corresponding pair of sub-codewords of length $q$, we focus on any given pair $\{\mathbf{X}^m\}_{m=1,2}$, where $\mathbf{X}^m=(X^m_1,\ldots,X^m_q)$. Thus, we use the subindex $i$ to denote the position of a given bit inside its corresponding sub-codeword. For the sake of simplicity we drop the super index $D$ from the set of received signals $\mathbf{Y}=\{\mathbf{Y}^{j,D}\}_{j=1,2,R}$ and rewrite the channel random variables for a given sub-codeword as $\mathbf{Y}_l^j=\{Y^j_{l,i}\}_{i=1}^q$.

In order to compute $p(\mathbf{y}_l|x_i^m)$, let us first focus on the \emph{a priori} information provided by the channel decoder. The sequence of random variables $\{X^m_i\}_{i=1}^q$ is assumed to be i.i.d. based on the fact that a spread interleaver is used to suppress strong dependencies (4-length cycles in the underlying factor graph) between the bits belonging to the same sub-codeword. Moreover, the p.m.f of its associated non-binary symbol $V^m=\psi_q(X^m_1,\ldots,X^m_q)$ is given by
\begin{equation}\label{eq:PV}
P_{V^m}(v) \triangleq\sum_{x^m_1,\ldots,x^m_q}\indicator{ v=\psi_q(x^m_1,\ldots,x^m_q)}\prod_{i=1}^{q} P^a_{X^m_i}(x^m_i),
\end{equation}
where the last factor represents the \emph{a priori} probabilities of $\{X^m_i\}_{i=1}^q$.

As shown in expression \eqref{eq:lc}, the non-binary symbols $V^1$ and $V^2$ are linearly combined over the finite field $GF(2^q)$, producing the non-binary symbol $V^R$. Therefore, the non-binary symbols joint p.m.f. can be factorized as
\begin{eqnarray}
P(v^1,v^2,v^R)=P(v^R\vert v^1,v^2)P(v^2\vert v^1)P(v^1)=P(v^1)P(v^2)\indicator{v^R=f(v^1,v^2)}, \label{eq:PV1V2VR}
\end{eqnarray}
where $V^R=f(V^1,V^2)$ is the linear combination defined in \eqref{eq:lc}. Now let us focus on the information coming from the channels. Due to the TDMA scheme, the following Markov chains are verified,
$\mathbf{Y}_l^1  \Leftrightarrow V^1 \Leftrightarrow (V^2, V^R, \mathbf{Y}_l^2, \mathbf{Y}_l^R)$;
$\mathbf{Y}_l^2  \Leftrightarrow V^2 \Leftrightarrow (V^1, V^R, \mathbf{Y}_l^1, \mathbf{Y}_l^R)$;
\mbox{$\mathbf{Y}_l^R  \Leftrightarrow V^R \Leftrightarrow (V^1, V^2, \mathbf{Y}_l^1, \mathbf{Y}_l^2)$}. Hence, we have
\begin{equation}\label{eq:PYV1V2VR}
p(\mathbf{y}_l\vert v^1,v^2,v^R)=p(\mathbf{y}_l^1\vert v^1)p(\mathbf{y}_l^2\vert v^2)p(\mathbf{y}_l^R\vert v^R).
\end{equation}

Furthermore, the non-binary symbols can also be expressed by the corresponding modulated symbols $\{S^j_i\}_{i=1}^q\in\{\pm 1\}^q$ as $V^j= \psi_q\left(\left\{(1+S_i^j)/2\right\}_{i=1}^q\right)\triangleq\theta_q\left(\{S_i^j\}_{i=1}^q \right)$, yielding
\begin{equation}\label{eq:PYjV}
p(\mathbf{y}_l^j\vert v^j)\propto\sum_{s^j_1,\ldots,s^j_q}\indicator{v^j=\theta_q(s^j_1,\ldots,s^j_q)}
\prod_{i=1}^q p(y^j_{l,i}\vert s^j_i),
\end{equation}
where the last product is due to the memoryless channel assumption made in this work, with
\begin{equation}
p(y_{l,i}^j\vert s_i^j)\propto \exp\left(-\frac{\left(y_{l,i}^j-\alpha^{j,D}\cdot \sqrt{d^{-\delta}_{j,D}}\cdot s_i^j\right)^2}{N_0}\right).
\end{equation}

Now, from expressions \eqref{eq:PV1V2VR} and \eqref{eq:PYV1V2VR} and by applying the Bayes theorem, we obtain the joint \emph{a posteriori} p.d.f. of the non-binary symbols as
\begin{equation}\label{eq:PV1V2VRY}
p(v^1,v^2,v^R\vert \mathbf{y}_l)\propto
p(\mathbf{y}_l^1\vert v^1)p(\mathbf{y}_l^2\vert v^2)p(\mathbf{y}_l^R\vert v^R)\cdot\indicator{v^R=f(v^1,v^2)}\frac{P^a(v^1)P^a(v^2)}{p(\mathbf{y}_l)}.
\end{equation}

Likewise, one can compute the \emph{a posteriori} p.m.f. $P(v^m\vert\mathbf{y}_l)$ of the non-binary symbol associated to a given source by marginalizing the previous equation. Thus,
\begin{equation}\label{eq:PVY}
P(v^m\vert \mathbf{y}_l)= p(\mathbf{y}_l^m\vert v^m)P^a(v^m)\cdot\sum_{v^{\overline{m}},v^R}\indicator{v^R=f(v^1,v^2)}
\frac{p(\mathbf{y}_l^{\overline{m}}\vert v^{\overline{m}})P^a(v^{\overline{m}})
p(\mathbf{y}_l^R\vert v^R)}{p(\mathbf{y}_l)}.
\end{equation}

On the other hand, following a similar reasoning as in \eqref{eq:PV}, we obtain
\begin{equation}
P(x^m_1,\ldots,x^m_q\vert\mathbf{y}_l)= \sum_{v^m}\indicator{v^m=\psi_q(x^m_1,\ldots,x^m_q)}P(v^m\vert\mathbf{y}_l),
\end{equation}
from which we can compute the bitwise channel conditional p.d.f. by marginalizing and applying again the Bayes Theorem, i.e.
\begin{equation}\label{eq:PYX1}
p(\mathbf{y}_l\vert x^m_i)=\sum_{\sim x_i^m,v^m}P(v^m\vert\mathbf{y}_l)\indicator{v^m=\psi_q(x^m_1,\ldots,x^m_q)}\frac{p(\mathbf{y}_l)}{P(x^m_i)},
\end{equation}

where $\sim x_i^m\triangleq \{x_j^m\}_{\forall j\neq i}$. Finally, combining \eqref{eq:PVY} and \eqref{eq:PYX1}, we get
\begin{equation}\label{eq:PYX2}
p(\mathbf{y}_l\vert x^m_i)=\sum_{\sim x_i^m,v^m}\indicator{v^k=\psi_q(x^m_1,\ldots,x^m_q)}\cdot p(\mathbf{y}_l^m\vert v^m)\prod_{i'\neq i} P^a(x^m_{i'})P^{\mbox{\tiny MARC}}(v^m),
\end{equation}
with
\begin{equation}\label{eq:P_MARC}
P^{\mbox{\tiny MARC}}(v^m)\triangleq\sum_{v^{\overline{m}},v^R}\indicator{v^R=f(v^1,v^2)}\cdot p(\mathbf{y}_l^{\overline{m}}\vert v^{\overline{m}})P^a(v^{\overline{m}}).
p(\mathbf{y}_l^R\vert v^R).
\end{equation}

\begin{figure}[!h]
\centering
  \includegraphics[width=\columnwidth]{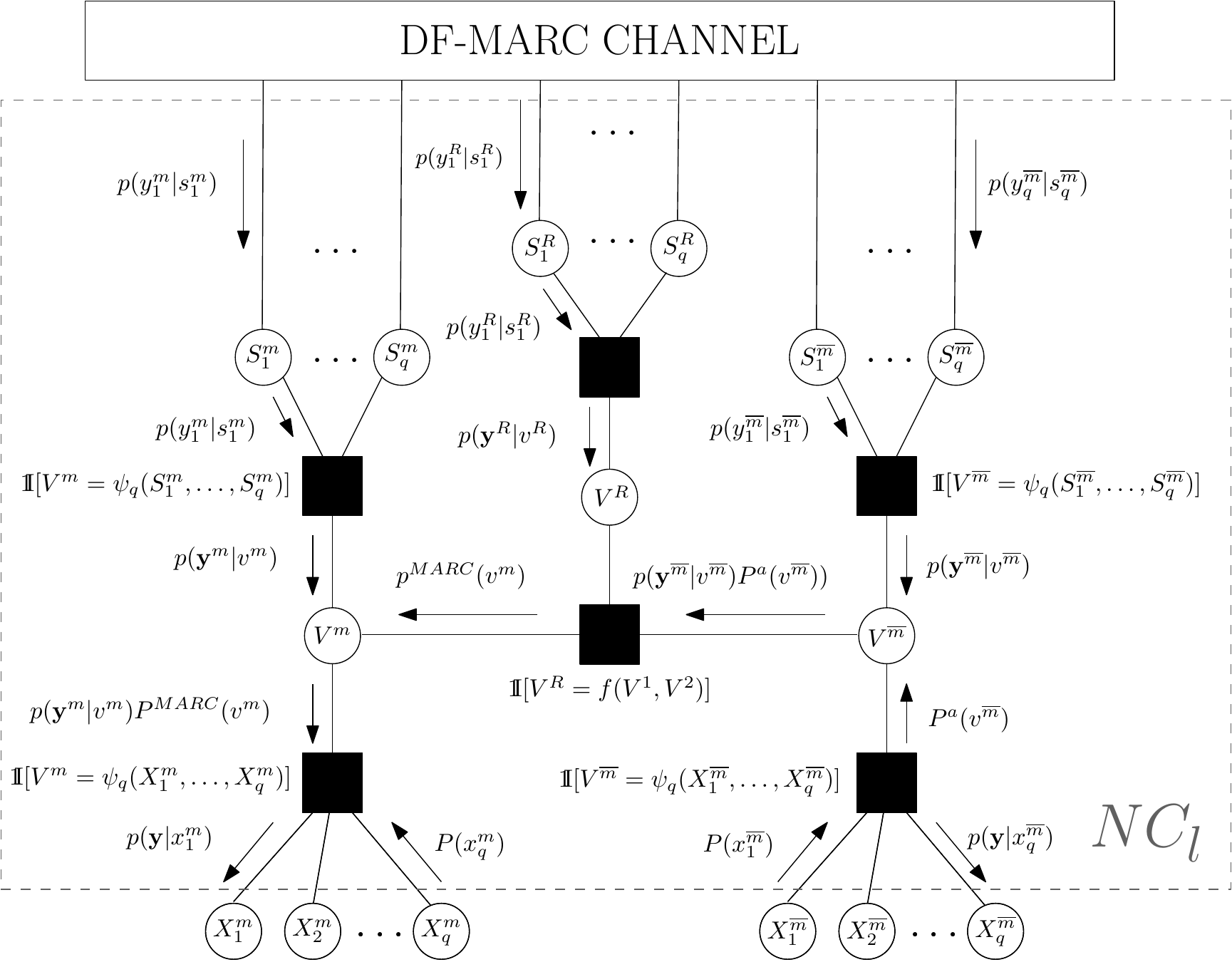}
  \caption{Factor Graph representing the factorization of $p(\mathbf{y}_l|x^m_i)$, i.e. the network check node $\mathcal{NC}_l$.}\label{fig:fNSD}
\end{figure}

The factorized form of $p(\mathbf{y}_l\vert x^m_i)$ given in expression \eqref{eq:PYX2} is graphically represented by the factor graph depicted in Figure \ref{fig:fNSD}, where for the sake of clarity the sub-index $l$ is dropped. The application of the SPA over the factor graph of Figure \ref{fig:fNSD} allows for a efficient computation of the likelihoods $\gamma(x_t^m)$. To be concise, the SPA iterates between the sub-factor graphs corresponding to the relay network code and the convolutional codes. It should be remarked that if the probability $p(\mathbf{y}_l^R|v^R)$ (message) is not dependent on the data from source $m$ (e.g., when the NC coefficients are set to zero or the relay-destination channel is in deep fade), then $P^{\mbox{\tiny MARC}}(v^m)$ will be uniformly distributed and consequently the exchange of messages between both convolutional decoders ($P^{\mbox{\tiny MARC}}(v^m)$) will not improve the performance of the decoder (see Fig. \ref{fig:fNSD}).


Although the factor graph shown in Fig. \ref{fig:decoder} has been constructed for convolutional codes, it could be easily modified if iteratively decodable codes (e.g. LDPC, Turbo) are used, by facing their outer codes with the Relay Network code subgraph. However, care should now be taken when programming the decoder activation scheduling.

In \cite{Brannstrom05} scheduling algorithms for both parallel and serially concatenated codes with several compounding graphs are proposed. We further refer to \cite{Brannstrom05} for algorithms that find the fastest convergent code activation schedule. Finally, a particular case  where non-binary LDPC codes were used as channel codes was published by current authors in \cite{Hernaez11}.

In the next section, the EXIT charts of the proposed JNCC are used to obtain a pair of NC-coefficients that optimizes the performance of the system. To simplify the exposition, the optimization procedure next presented assumes that both sources are detected with no errors at the relay, in line with the assumptions made in the related literature \cite{Hausl05,Nguyen07,Chebli09,Hernaez11}.

\section{Analysis of the Relay Network Code based on EXIT charts} \label{sec:EXIT}

An additional insight can be gained through an analysis of the NC-coefficient based on the interchange of mutual informations between the output of the channel decoders and the output of the relay network decoder. Due to the iterative nature of the decoding algorithm, EXtrinsic Information Transfer (EXIT, see \cite{EXIT}) charts are a good method for visually exploring this iterative exchange of information. Given a code, the EXIT function associated is defined by the relation between the \emph{a priori} mutual information at the input of the decoder (commonly denoted as $I_a$) and the corresponding extrinsic mutual information $I_e$ at its output, i.e. $I_e=T(I_a)$. For further information on EXIT charts we refer \cite{EXIT} and references therein to the reader.

We will denote the transfer function of the network code for a given $q$ and $\mathbf{h}$ as $I^{\mbox{\tiny NC} }_e=T_{\mathbf{h}}^q(I^{\mbox{\tiny NC} }_a)$. Notice that, $T_{\mathbf{h}}^q(0)$ and $T_{\mathbf{h}}^q(1)$ represent the extrinsic information at the output of the network decoder with no \emph{a priori} and full \emph{a priori} information about the information bits, respectively. Moreover, since the mutual information at the input of the channel decoders is equal to the mutual information at the output of the network decoder, i.e. $I_a^{\mbox{\tiny CC}}=I_e^{\mbox{\tiny NC} }$, the extrinsic mutual information at the output of the convolutional decoders is given by $I_e^{\mbox{\tiny CC} }=T^{\mbox{\tiny CC} }(I_e^{\mbox{\tiny NC} })$. Thus, for a successful decoding procedure, there must be an open gap between both EXIT curves so that the iterative decoding can proceed from $I_e^{\mbox{\tiny CC} }=0$ to $I_e^{\mbox{\tiny CC} }= 1$\footnote{Although for a perfectly successful decoding the final $I_e^{\mbox{\tiny CC} }$ should be equal to one, we also consider the values of $I_e^{\mbox{\tiny CC} }\approx 1$ that yield to negligible error floors.}. When both transfer functions intersect, the iterative process will stop at a given extrinsic mutual information of the source bits $I_e^{\mbox{\tiny CC} }< 1$. A crossing yielding $I_e^{\mbox{\tiny CC} }\leq 0.5$ will be referred to as early-crossing and as late-crossing, otherwise (i.e., $0.5<I_e^{\mbox{\tiny CC} }<1$). Since the transfer functions are monotonically increasing, the higher the value $T_{\mathbf{h}}^q(0)$ is, the later the early-crossing will occur. On the other hand, and if no early-crossing occurs, the higher the value of $T_{\mathbf{h}}^q(1)$ is, the closer $I_e^{\mbox{\tiny CC} }$ will be to one.

Next, we justify the reason why the NC coefficients have a greater impact on the value of $T_{\mathbf{h}}^q(1)$ than on $T_{\mathbf{h}}^q(0)$. When no \emph{a priori} information is available at the network decoder, all its information comes from the sources- and relay-destination links. On the other hand, when full \emph{a priori} information is available, only the information provided by the relay-destination link is relevant, since the information regarding the coded messages from the sources is fully supplied by the channel decoders. In addition, from the Area Theorem of EXIT charts \cite{EXIT}, which states that the area below the transfer function depends only on the rate of the encoder, the area below the transfer functions will be constant $\forall q, \mathbf{h}$. Consequently, the transfer functions of those NC-coefficients with a large value of $T_{\mathbf{h}}^q(1)$ are expected to be flat shaped at low values of $I_e^{\mbox{\tiny NC} }$, and steep shaped at values near one.

\begin{figure}[!t]
  \centering
  \includegraphics[width=\columnwidth]{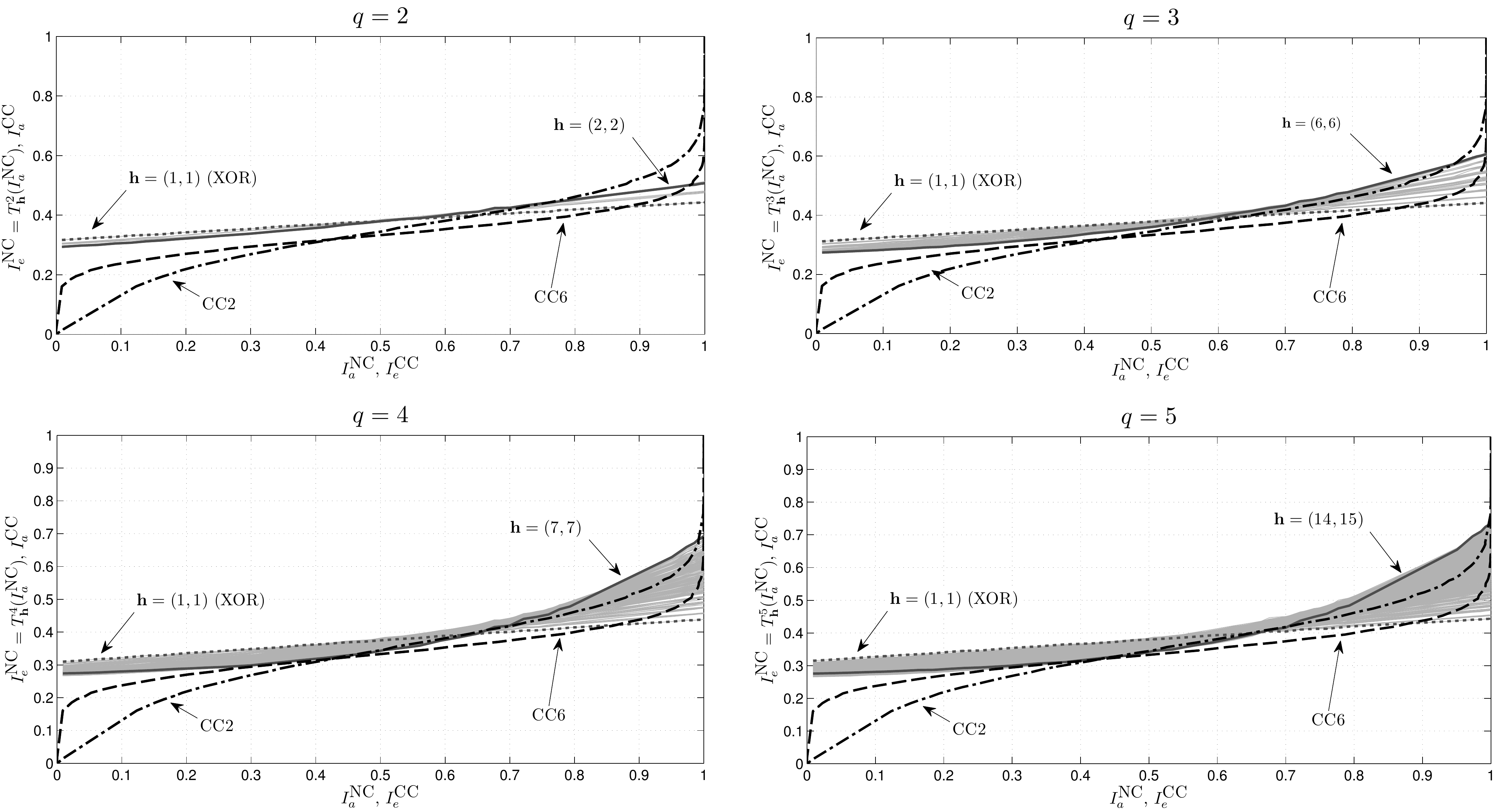}
  \caption{EXIT charts for different network decoders and convolutional codes.}\label{fig:EXIT}
\end{figure}

To corroborate the above, let us first assume AWGN channels (i.e. $\bm{\alpha}=(1,1,1)$). Figure \ref{fig:EXIT} plots the EXIT chart of a network check node for different NC-coefficients and values of $q$, along with the transfer function of a $2^2$-state (CC2) and a $2^6$-state (CC6) convolutional code with same transmission rate $1/3$. Note that by the symmetry of the network, it is sufficient to consider only those NC-coefficients $\mathbf{h}=(\rho_i,\rho_j)$ with indexes verifying $1\leq i\leq j\leq 2^q-1$. For the sake of clarity, we will denote $\mathbf{h}=(\rho_i,\rho_j)$ by $(i,j)$. From this figure, it can be observed that the transfer function corresponding to $\mathbf{h}=(1,1)$ (i.e. raw XOR-network coding) maximizes $T_{\mathbf{h}}^q(0)$ and minimizes $T_{\mathbf{h}}^q(1)$ for all $q$. Also observe that for high values of $q$ and for some NC coefficients, the initial flat shape of the plotted transfer functions curves induces an early-crossing with the curve of CC2, and consequently the system requires more energy per symbol to open the gap between these curves. In this case, one may opt to select NC-coefficients that increase the value of $T_{\mathbf{h}}^q(0)$, so the early-crossing could be avoided. As a drawback, a higher error floor due to their associated lower $T_{\mathbf{h}}^q(1)$ is expected. As highlighted in the figures, such alternate coefficients are given by $\mathbf{h}=(2,2)$ ($q=2$), $\mathbf{h}=(6,6)$ ($q=3$), $\mathbf{h}=(7,7)$ ($q=4$) and $\mathbf{h}=(14,15)$ ($q=5$).

Therefore, for AWGN channels we can conclude that by a proper choice of the coefficients an iterative gain will be achieved when using a linear combination of the estimated symbols, as it was also stated in \cite{Yang07}. This contrasts with the more commonly used scheme based on a channel encoder at the relay. Moreover, from the variety of transfer functions generated by the family of NC coefficients, it is also concluded that a tailored non-binary linear combination at the relay can outperform 1) the XOR coding method (i.e. $\mathbf{h}=(1,1)$) first proposed in \cite{yChen06,Yang07,Yu10}; and 2) a random choice of such parameters as was proposed in \cite{Xiao09,Guo09}.

Before considering the more general case of fading channels, observe that the proposed soft-output network decoder can be seen as $N/q$ parallel systematic codes $\mathcal{W}(\mathbf{h})$; each of them having the $3q$-length codewords $\mathbf{W}=(\{X^1_i\}^q_{i=1},\{X^2_i\}^q_{i=1},\{X^R_i\}^q_{i=1})$ where its systematic bits, $(\{X^1_i\}^q_{i=1},\{X^2_i\}^q_{i=1})$, correspond to the interleaved bits of the sub-sequences from sources $\mathcal{S}^1$ and $\mathcal{S}^2$, respectively, and its parity bits, $\{X^R_i\}^q_{i=1}$, correspond to the bits generated by the relay. Hence, each pair of NC-coefficients generates a particular code $\mathcal{W}(\mathbf{h})$ with a different distance spectrum and therefore, different transfer functions. Moreover, the relation among these functions depends solely on their distance spectrum and it is independent of the quality of the channels.

\begin{figure}[!ht]
  \centering
  \includegraphics[width=\columnwidth]{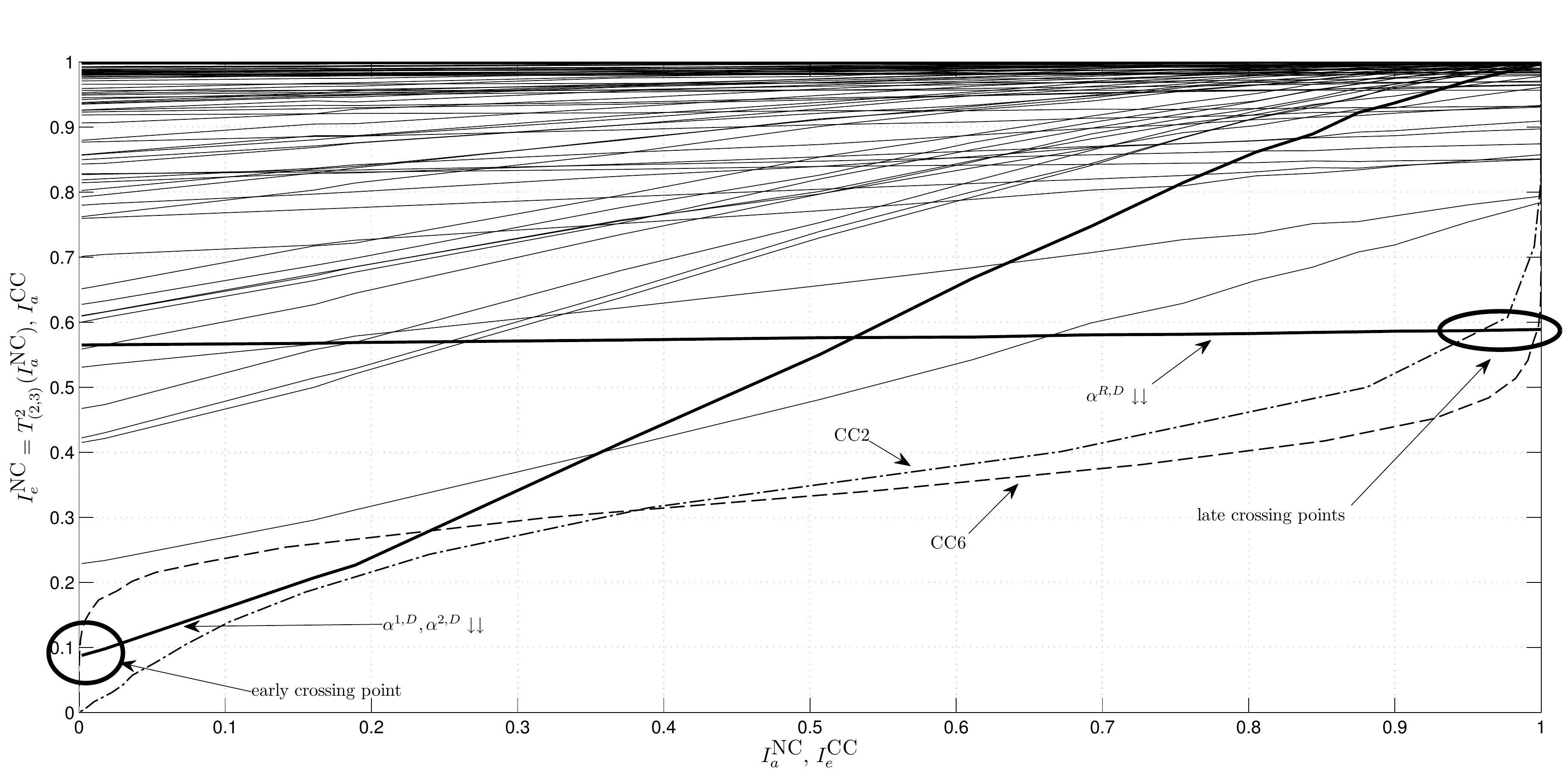}
  \caption{EXIT curves of 100 system usages for $q=2$ and $\mathbf{h}=(2,3)$.}\label{fig:EXIT_fade}
\end{figure}

We now look at Rayleigh fading links. In this case, different realization of the coefficients of the Rayleigh fading $\bm{\alpha}$ will produce differently shaped transfer functions, making $T_{\mathbf{h}}^q$ a random mapping from $\mathbb{R}\rightarrow \mathbb{R}$. If the channel between the relay and the destination is in a deep fade, no gain is obtained by iterating (since $P^{\mbox{\tiny MARC}}(\cdot)$ is uniform), and a late-crossing might occur (given that $T_{\mathbf{h}}^q(0)$ is high enough). On the other hand, if a deep fade occurs in both source-destination channels, an early-crossing could be produced due to the low value of $T_{\mathbf{h}}^q(0)$. Both situations are shown in Figure \ref{fig:EXIT_fade}, which plots the EXIT curves of $100$ channel realizations for $q=2$ and $\mathbf{h}=(2,3)$. In this Figure, the curves corresponding to two fading realizations producing an early- and late-crossings, have been highlighted for clarity. Since a decoding error is produced when a crossing occurs, the probability of a failed decoding event could be approximated by $\mathcal{P}\{\mbox{failed decoding}\}\approx \mathcal{P}\{\mbox{early-crossing}\}+ \mathcal{P}\{\mbox{late-crossing}\}$.

It should be mentioned that the $\mathcal{P}\{\mbox{early-crossing}\}$ strongly depends on the quality of the channels whereas $\mathcal{P}\{\mbox{late-crossing}\}$ is not so dependant. The reason being that increasing the received signal-to-noise ratio will reduce the probability of an early-crossing, since the decoder would be fed during the first iteration with more reliable channel information. However, the late-crossing probability will not be significantly reduced since the influence of the information provided by both source-destination channels decreases as the channel decoders begin to provide \emph{a priori} information regarding the source encoded bits. Therefore, as the signal-to-noise ratio increases, the late-crossing probability starts to dominate the failure probability, and hence, beyond a certain value of the signal-to-noise ratio, the early-crossing probability becomes negligible, regardless of the channel code used.

In addition, the failure probability will also depend on the type of convolutional code used and as shown in Fig. \ref{fig:EXIT_fade}, the more complex the code is, the larger the value of $T_{\mathbf{h}}^q(0)$ should be in order to avoid early-crossings. On the contrary, the less complex the code is, the higher the value of $T_{\mathbf{h}}^q(1)$ should be in order to avoid late-crossings. Therefore, the value of $T_{\mathbf{h}}^q(0)$ and $T_{\mathbf{h}}^q(1)$ will strongly determine the failure probability of the proposed decoder. To analyze the crossing probabilities one has to statistically characterize the behavior of the random variables $T_{\mathbf{h}}^q(0)$ and $T_{\mathbf{h}}^q(1)$ for different values of $q$ and NC-coefficients. To that end, we next show that the distribution of these random variables depend on the path-loss gain suffered by the signal coming from the relay (i.e. the position of the relay).

For $T_{\mathbf{h}}^q(1)$ (i.e., the full \emph{a priori} case) the extrinsic information generated by the network check nodes (i.e., $P^{\mbox{\tiny MARC}}(\cdot)$) depends on the quality of the relay-destination channel through the check node associated to the linear combination. Therefore, in the limit when the SNR of the relay-destination channel tends to infinity, the probability of $\mathcal{P}(T_{\mathbf{h}}^q(1)=1) $ tends to one. As a consequence, the slope of transfer functions will increase since $T_{\mathbf{h}}^q(0)$ remains fairly constant; and consequently, some of the early-crossings will be avoided.

\begin{figure}[!ht]
\centering
  \includegraphics[width=\columnwidth]{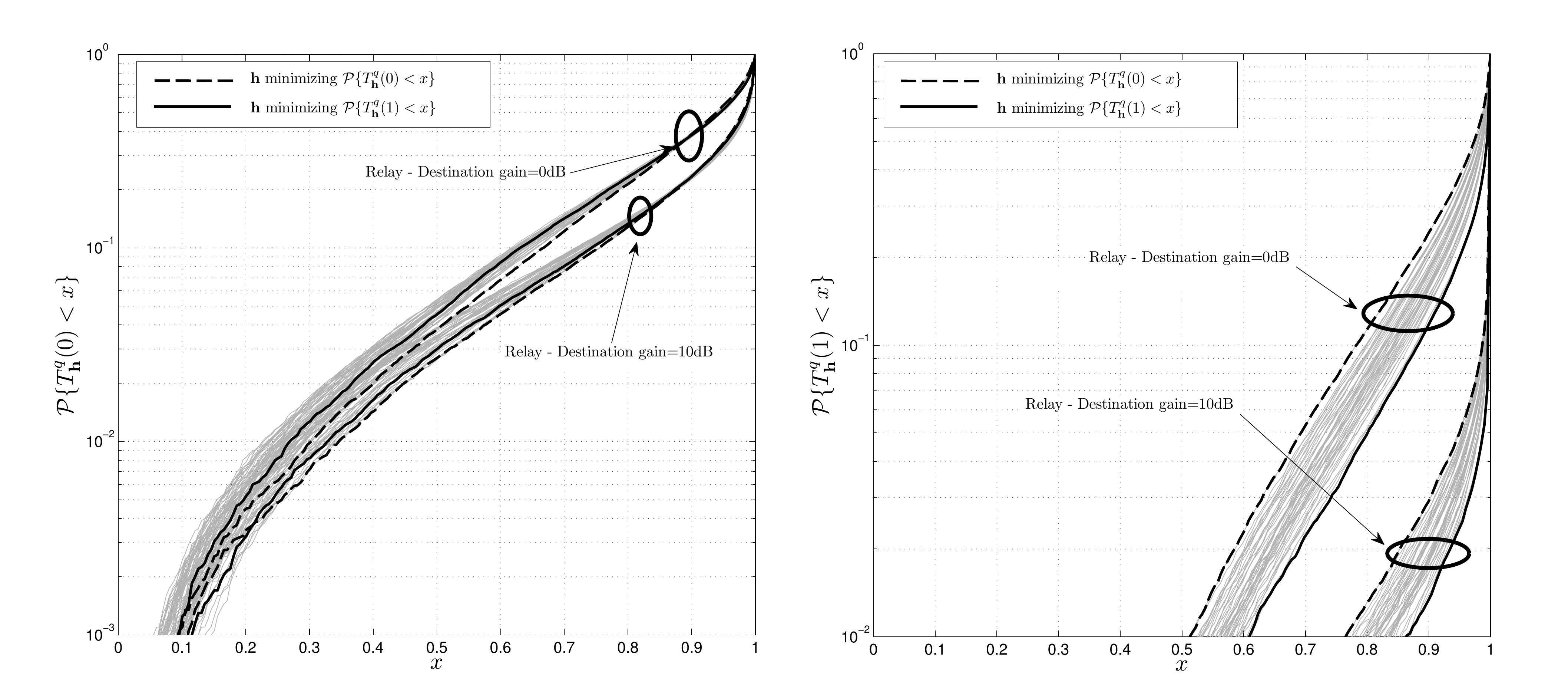}
  \caption{CDF of $T_{\mathbf{h}}^q(0)$ and $T_{\mathbf{h}}^q(1)$ for $q=3$ and all possible values of $\mathbf{h}$.}\label{fig:cdf_comp}
\end{figure}

To make this analysis clearer, Figure \ref{fig:cdf_comp} shows the Cumulative Distribution Function (CDF) of $T_{\mathbf{h}}^q(0)$ and $T_{\mathbf{h}}^q(1)$ for $q=3$ and all combinations of $\mathbf{h}$ for the cases where: the relay and the sources are at the same distance to the destination (0 dB gain), and the relay is placed at half distance between the sources and destination ($10$ dB gain). As mentioned for the AWGN case, it can be observed that the choice of a different value of $\mathbf{h}$ has a stronger impact on $T_{\mathbf{h}}^q(1)$ than on ($T_{\mathbf{h}}^q(0)$. Also, increasing the quality of the relay-destination channel leads to an increase in the values of the realizations of $T_{\mathbf{h}}^q(1)$, which in turn diminishes $\mathcal{P}\{T_{\mathbf{h}}^q(1)<x\}$ for any $x\in [0,1]$. Moreover, $\mathcal{P}\{T_{\mathbf{h}}^q(0)<x \}$ does not significantly change as the gain in the relay-destination link is varied.

Observe from Figure \ref{fig:cdf_comp} that given a $q$, there is a particular value $\mathbf{h}^*_1$ (shown as solid lines) that minimizes\footnote{ The minimization is performed by an exhaustive search over all possible $\mathbf{h}\in GF(2^q)$ (Figure \ref{fig:cdf_comp}).} $\mathcal{P}\{T_\mathbf{h}^q(1)<x \}$, $\forall x\in [0,1]$ regardless the quality of the channels. Similarly, there is a value $\mathbf{h}^*_0$ (shown as dashed lines) that minimizes $\mathcal{P}\{T_\mathbf{h}^q(0)<x\}$, $\forall x\in [0,1]$.

At this point, by denoting the smaller SNR value that makes the term $\mathcal{P}\{\mbox{late-crossing}\}$ to dominate the system performance as $SNR_{th}$, the best choice of NC-coefficients is $\mathbf{h}^*_1$ when $SNR\geq SNR_{th}$, and $\mathbf{h}^*_0$, otherwise. Based on this, the steps for selecting the NC-coefficients can be summarized as follows: 1) find the values of $\mathbf{h}^{*}_0$ and $\mathbf{h}^{*}_1$; 2) estimate by Monte Carlo simulations the value of $SNR_{th}$; and 3) choose $\mathbf{h}^{*}_0$ or $\mathbf{h}^{*}_1$ depending on the operating region of the system.

In conclusion, the value of a good $\mathbf{h}$ will depend on the channel codes, the relay position and channel conditions. This corroborates the previous statement that a random choice of the NC-coefficient might lead to some performance degradation.

Finally, note that if the channel codes are composed of several subgraphs (e.g. LDPC or Turbo codes), several transfer functions (one for each compounding subgraph) are obtained and a direct representation of these functions will result in an $N$-dimensional EXIT chart. By converting the $N$-dimensional EXIT chart into a two-dimensional EXIT chart, for example by using the  EXIT Chart Projection Algorithm proposed in \cite{Brannstrom05}, the above analysis could still be applied.

\section{Simulation Results}\label{sec:res}

In order to asses the performance of the proposed scheme and to corroborate the conclusions from the previous analysis, several sets of simulations have been performed. We have considered a symmetric scenario where both source nodes are placed at the same distance from the destination nodes, i.e. $d_{1,D}=d_{2,D}$. Regarding the relay position, three different scenarios are defined:
\begin{itemize}
\item[A)] The relay and the sources are deployed at identical distance from the destination, (i.e.,  $SNR_{R,D}=SNR_{1,D}+0$dB). This setup was used in \cite{Hausl05,Hausl06}.

\item[B)] The distance between the relay and the destination is three quarters of the source-destination distance ($SNR_{R,D}=SNR_{1,D}+4.4$dB) as used in \cite{Chebli09,Hernaez11}.

\item[C)] The relay is set at approximately half the distance between the sources and the destination ($SNR_{R,D}=SNR_{1,D}+10$dB). This setup was also considered in \cite{Hausl06}.
\end{itemize}

The channel codes used at both source nodes are identical, non-systematic convolutional codes of rate 1/3.  Two types of convolutional codes are considered: a $2^2$-state  $[5, 7, 7]_8$ code\footnote{The subindex $8$ in the definition of the code stands for \emph{octal}.} (heretofore denoted as CC2), and a $2^6$-state $[554, 624, 764]_8$ code (correspondingly, CC6). A zero-bit termination tail is appended at each source sequence. We use packets of $K=32$ bits (i.e. we use $M=32\cdot 3/2=48$), and the interleavers have been randomly generated and are independent of each other. We have considered 4-QAM\footnote{Further improvement can be expected with higher order modulation schemes and by applying the so-called Bit-Interleaved Coded Modulation (BICM \cite{Biglieri98}) technique. However, this research line lies beyond the scope of this manuscript.}, which leads to an spectral efficiency of $\rho=4/9$ [bits per complex dimension], and the number of iteration for the SPA has been set to $\mathcal{I}=15$. Finally, from the region of achievable decode-and-forward rates of the TD-MARC given in \cite{Hausl_thesis}, (derived from the capacity bounds of the C-MARC in \cite{Sankar04}), a set of upper bounds on the outage probabilities has been obtained by specifying the actual packet lengths of the source information bits and the transmitted sequences ($K=32$, $M=48$). These outage probability bounds are used as an information-theoretic benchmark for the Packet Error Rate (PER), where a packet is in error if one or both sources packets are erroneously decoded.

\begin{figure}[!ht]
\centering
  \includegraphics[width=\columnwidth]{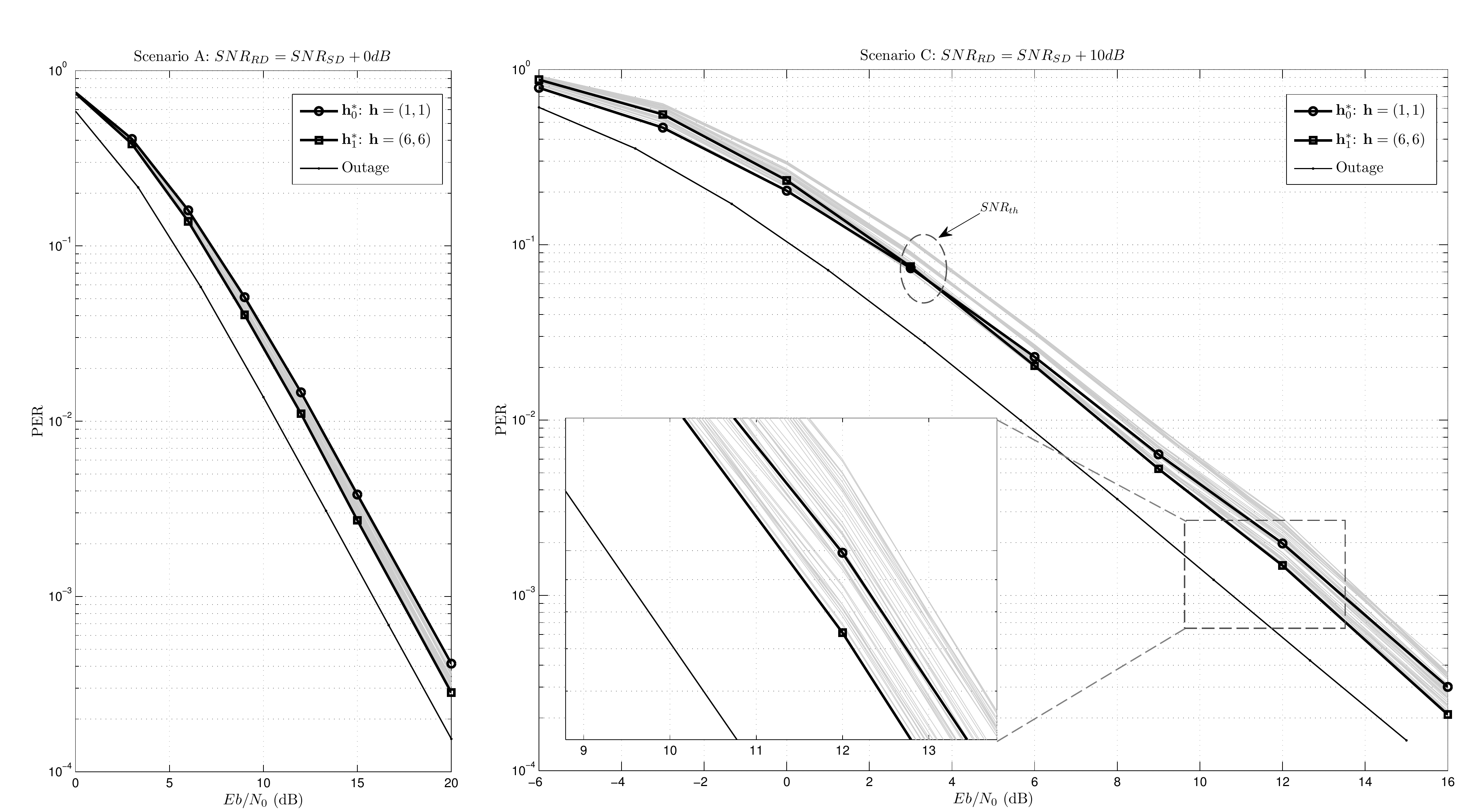}\\
  \caption{Outage probabilities and PER performance of the proposed approach when the CC2 convolutional code is used.}
  \label{fig:CC2}
\end{figure}

\begin{figure}[!ht]
\centering
  \includegraphics[width=\columnwidth]{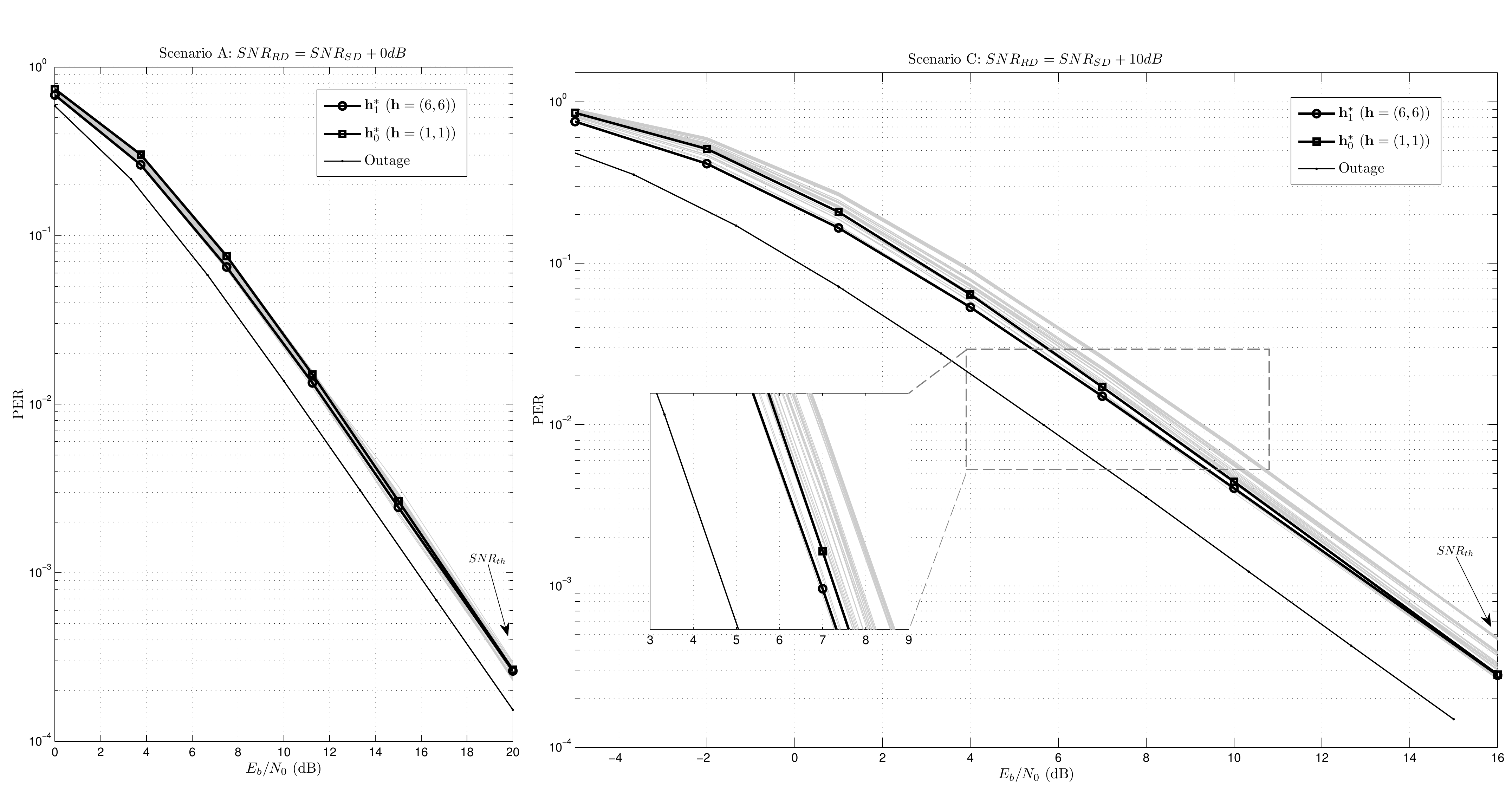}\\
  \caption{Outage probabilities and PER performance of the proposed approach when the CC6 convolutional code is used in scenario A (left) and C (right).}\label{fig:q7}
\end{figure}

Based on the analysis carried out in Section \ref{sec:EXIT}, the NC-coefficients for $q=3$ that minimizes $\mathcal{P}\{T_\mathbf{h}^q(1)<x \}$ and $\mathcal{P}\{T_\mathbf{h}^q(0)<x \}$ are given by $\mathbf{h}=(6,6)$ and $\mathbf{h}=(1,1)$, respectively.
To corroborate the optimality of these values a first set of simulations have been done for scenarios A and C (Fig. \ref{fig:CC2}-\ref{fig:q7}). Since we are mainly interested in the selection of the NC coefficients, in these simulations error-free links between the sources and the relay are assumed. Nevertheless, the noisy source-relay link case is also discussed later in this section.

By using the channel code CC2, Figure \ref{fig:CC2} plots the end-to-end PER versus $E_b/N_0=SNR-10\log_{10} \rho$ (in dB) for all possible values of $\mathbf{h}$ and for scenarios A (lefthand plot) and C (righthand plot). It can be observed that in scenario A, and for all values of SNR, the coefficients $\mathbf{h}=(6,6)$ (i.e. $\mathbf{h}^*_1$) are the optimal choice. For scenario C, $\mathbf{h}^*_1$ is still optimal in the range $SNR>SNR_{th}$ (recall that $SNR_{th}$ denotes the $SNR$ at the crossing of the curves for $\mathbf{h}^*_1$ and $\mathbf{h}^*_0$); however, for $SNR<SNR_{th}$ the optimal choice is $\mathbf{h}=(1,1)$ (i.e. $\mathbf{h}^*_0$). This corroborates what was stated in Section \ref{sec:EXIT}, that the use of low-complexity codes reduces the influence of $\mathcal{P} \{\mbox{early-crossing}\}$, leading to low values of $SNR_{th}$ (in scenario A $SNR_{th}<0$ and in scenario C $SNR_{th}\approx 3$). Moreover, it can be seen that as the quality of the relay-destination link increases (going from scenario A to C), the value of $SNR_{th}$ increases; as also stated in Section \ref{sec:EXIT}.

Figure \ref{fig:q7} shows a similar analysis on the PER when using the CC6 code, instead. It can be observed that in both scenarios $SNR_{th}$ arise at high values of SNR, since $\mathbf{h}^*_0$ outperforms $\mathbf{h}^*_1$ for the SNR ranges of interest (in scenario A and C, the crossing  occurs at $SNR>20$dB and $SNR>16$dB, respectively), corroborating the fact that we are using high-complexity convolutional codes (see Section \ref{sec:EXIT}).

Furthermore, these results show that as the relay is placed toward the destination, a tailored selection of the NC-coefficients improves the performance of the system; e.g. in scenario C, gains of 1.3dB and 1.5dB, with respect to the worst selection, can be achieved when using CC2 and CC6, respectively . One can also conclude from the figures that, since the performances for the different coefficients are roughly uniformly distributed between the best and worst case scenarios, a gain of 0.75dB (scenario C and CC6) is obtained with respect to the random choice of coefficients.

The next set of simulation compares the gap between the performance of the proposed system with respect to the outage probability. This gap is also computed for different schemes found in the literature, all having a spectral efficiency less than 2 (bits per complex dimension). It should be mentioned that all these schemes \cite{Hausl05,Hausl06,Chebli09,Hernaez11} were analyzed assuming error free source-relay channels. Therefore, the outage probabilities (upper bounds) for scenarios A,B,C has been computed with the later assumption and for unconstrained channel inputs (Gaussian).

These gaps are given in Table \ref{table:gap}, where the first two columns refer to our scheme.
Observe that the proposed scheme outperforms all the reference schemes. Moreover, the proposed system achieves these results by using short-length codewords: 144 complex dimensions per use of TD-DF-MARC in contrast to the 2176 utilized in \cite{Chebli09}, the 6000 utilized in \cite{Hausl05,Hausl06} or the 27000 utilized in \cite{Hernaez11}.

\begin{table}[!ht]
\renewcommand{\arraystretch}{1.3}
\caption{Gaps to the outage probabilities}
\label{table:gap}
\centering
\begin{tabular}{|c|c|c|c|c|c|c|}
\hline
\bfseries Scenarios & \bfseries $\textbf{q=3}$, CC2 & \bfseries $\textbf{q=1}$, CC6 & \cite{Hernaez11} & \cite{Hausl05} & \cite{Hausl06} & \cite{Chebli09}\\
\hline\hline
{Scenario A} & {1.39 dB} & {1.44 dB} & {-} & 3.4 dB & 2.7 dB & - \\
\hline
{Scenario B} & {1.64 dB} & {1.64 dB} & {1.7 dB} & - & - & 4.8 dB  \\
\hline
{Scenario C} & {2.04 dB} & {2.36 dB} & {-} & - &  5.2 dB & -  \\
\hline
\end{tabular}
\end{table}

\begin{figure}[ht]
\centering
  \includegraphics[width=\columnwidth]{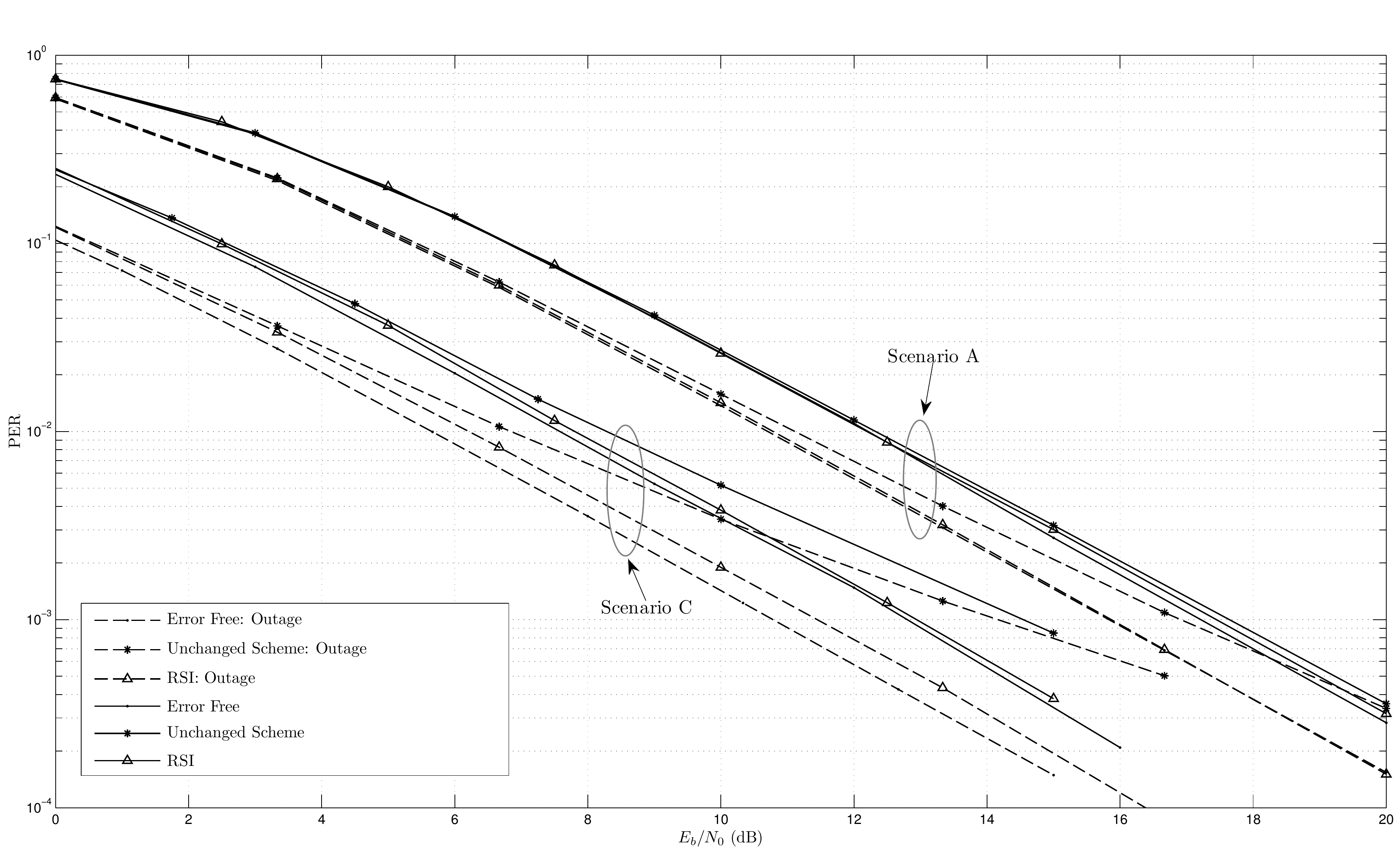}\\
  \caption{PER performance of the proposed schemes for the erroneous source-relay channel setup when the CC2 convolutional code and $\mathbf{h}=(6,6)$ is used.}
  \label{fig:EP}
\end{figure}

Finally, we consider the case where non-ideal source-relay channels are used. An unchanged implementation of the proposed scheme will lead to error propagation at the decoder and as a result a degradation on the performance as shown in Figure \ref{fig:EP}. Moreover, at high SNRs, the closer the relay to the destination is, the stronger the impact of the error propagation will be. However, since the error propagation also degrades the outage probability as shown in Fig. \ref{fig:EP}, the gap between the performance of the unchanged scheme and the new outage probability curve\cite{Hausl_thesis} is still small, and is in fact negligible at high SNRs. We conclude that our scheme is still robust when non ideal channel error propagation occurs.

An alternative scheme to cope with the error propagation at the destination is as follows. This scheme assumes that the destination node knows if an error has occurred at the relay (Relay State Information, RSI), e.g, by using an additional low rate error free relay-destination channel. To this end, a cyclic redundancy check (CRC) is added at the end of the source transmitted sequences, so that the relay can detect if any residual errors have occurred in $\widehat{\mathbf{X}}^1$ and $\widehat{\mathbf{X}}^2$. If errors are detected in $\widehat{\mathbf{X}}^m$, the relay sets $h^m$ to $0$ (i.e. the sequence transmitted by the relay only conveys information from $\mathcal{S}^{3-i}$) and the relay communicates that an error has been produced to the destination. The decoder sets at the network check node, $h^m$ to $0$, avoiding in this way error propagation through channel decoders. However, some performance degradation is still expected since $P^{\mbox{\tiny MARC}}(v^m)$ (see Fig. \ref{fig:fNSD}) is now uniformly distributed and therefore no iteration gain will be obtained.  The performance of the RSI scheme is shown in Figure \ref{fig:EP} together with the corresponding outage probability. One can observe that the performance loss with respect to the error-free case is negligible.

\section{Conclusions} \label{sec:con}
This paper proposes a novel joint network-channel coding scheme for the Decode-and-Forward Time-Division Multiple Access Relay Channel. Specifically, we have designed a Joint-Network-Channel code which does not perform channel coding on the already network-coded bits, reducing the complexity at the relay node without compromising performance. A method for selecting the best pair of coefficients of the linear combination is derived based on an EXIT charts analysis. Moreover, the proposed code allows the sources to use completely different channel codes, at the sole expense of an increased complexity on the EXIT chart analysis. The decoding at the destination is performed by applying the SPA over the derived factor graph of the JNCC code. Monte Carlo simulations show that the proposed scheme outperforms, in terms of its gap to the corresponding outage probability upper bounds, the previously published schemes for the same network setup. Besides, this gain is achieved by using short-length codewords, which makes our proposal particularly attractive for low-latency applications.

\section*{Acknowledgments}
The authors would like to thank the Spanish Ministry of Science \& Innovation for its support through the \emph{COMONSENS} (CSD200800010) and \emph{COSIMA} (TEC2010-19545-C04-02) projects.

\bibliographystyle{IEEEtran}

\bibliography{./biblio_JSAC_short}

\end{document}